# Microbial interactions in marine water amended by eroded benthic biofilm: A case study from an intertidal mudflat

Hélène Montanié[1*], Pascaline Ory[1], Francis Orvain[2], Daniel Delmas[3], Christine Dupuy[1], Hans J. Hartmann[1]

[1] Littoral, Environnement et SociétéS (LIENSs), Université de La Rochelle, UMR 7266 CNRS-ULR, 2 rue Olympe de Gouges, 17000 La Rochelle Cedex, France

[2] Université de Caen Basse-Normandie, FRE3484 BioMEA CNRS, Esplanade de la Paix, 14032 Caen, France

[3] IFREMER, Dyneco, Laboratoire Pelagos, BP70, 29280 Plouzané, France

* Corresponding author : Hélène Montanié: Littoral, Environnement et SociétéS (LIENSs), Université de La Rochelle, UMR 7266, 2 rue Olympe de Gouges, 17000 La Rochelle Cedex, France email: helene.montanie@univ-lr.fr tel: +33546458280

**Abstract**

In shallow macrotidal ecosystems with large intertidal mudflats, the sediment-water coupling plays a crucial role in structuring the pelagic microbial food web functioning, since inorganic and organic matter and microbial components (viruses and microbes) of the microphytobenthic biofilm can be suspended toward the water column. Two experimental bioassays were conducted in March and July 2008 to investigate the importance of biofilm input for the pelagic Microbial and Viral Loops. Pelagic inocula (< 0.6µ- and < 10µ filtrates) were diluted either with <30kDa-ultrafiltered seawater or with this ultrafiltrate enriched with the respective size-fractionated benthic biofilm or with <30kDa-benthic compounds (BC). The kinetics of heterotrophic nanoflagellates (HNF), bacteria and viruses were assessed together with bacterial and viral genomic fingerprints, bacterial enzymatic activities and viral life strategies. The experimental design allowed us to evaluate the effect of BC modulated by those of benthic size-fractionated microorganisms (virus+bacteria, +HNF). BC presented (1) in March, a positive effect on viruses and bacteria weakened by pelagic HNF. Benthic microorganisms consolidated this negative effect and sustained the viral production together with a relatively diverse and uneven bacterial assemblage structure; (2) in July, no direct impact on viruses but a positive effect on bacteria modulated by HNF, which indirectly enhanced viral multiplication. Both effects were intensified by benthic microorganisms and bacterial assemblage structure became more even. HNF indirectly profited from BC more in March than in July. The Microbial Loop would be stimulated by biofilm during periods of high resources (March) and the Viral Loop during periods of depleted resources (July).

# 1 Introduction

In shallow estuarine ecosystems, hydrodynamics (tide, waves, wind) induce erosion of particulate material from the sediments into the overlying water column and at the sediment-water interface, the gradients in organic and inorganic nutrients may structure the composition of the microbial communities and their activities (Seymour et al., 2007). Differences in grain size distribution may explain the distribution of dissolved and particulate matter according to the porosity and permeability of the sediment (Pinto et al., 2013). The smaller pore size and higher surface area in clays and silts compared to sand may infer a higher dissolved nutrients in clayed sediment and a higher desorption of organic particles in sandy sediment (Wainright, 1987). Consequently, understanding the pelagic microbial food web dynamics of an intertidal muddy ecosystem requires consideration of the benthic-pelagic coupling. Sediment suspension may be an efficient way to provide inorganic nutrients for phytoplankton and to deliver sedimentary organic matter (Particulate Organic Matter and Dissolved Organic Matter) for pelagic heterotrophic entities (Arfi and Bouvy, 1995; Garsteki et al., 2002; Hopkinson et al., 1998; Wainright, 1987, 1990).

The resuspended benthic organic matter, used by bacteria, is channeled to the higher trophic levels via the heterotrophic nanoflagellate (HNF) bacterivory within the pelagic microbial loop (Azam et al., 1983), while viral lysis tends to maintain the matter at lower levels by shunting those fluxes (Bratbak et al., 1992; Wilhelm and Suttle, 1999). Microorganisms may also be resuspended (Dupuy et al., this issue; Shimeta et al., 2002). In the water column, the dynamics of benthic viruses may be globally governed by the trade-off between their irreversible and non-infective adsorption onto benthic-suspended matter (Suttle and Chen, 1992) and their multiplication at the expense of more active attached bacteria (Kernegger et al., 2009; Riemann and Grossart, 2008). The abundance of HNF depends directly on the resuspension of their benthic representatives and would also, in the case of muddy sediments, indirectly benefit from the increase of their prey (bacteria and pico-, nanoautotrophs), which benefit from the input of benthic nutrients (Garstecki et al., 2002; Guizien et al., 2013). Thus, the microbial interactions include HNF bacterivory, HNF virivory (Bettarel et al., 2005), viral lysis of bacteria, HNF or viruses and coincidental HNF predation of infected bacteria (Miki and Jacquet, 2008). Apart from these direct links, the availability of nutritive resources, including the feed-back effect of the regenerated products from lysis and predation, and the presence/absence of the respective second member of the predator guild of bacteria (virus, HNF) may influence the expression of the bacterial resistance to predators and the life cycles of temperate viruses (Miki and Jacquet, 2008, 2010; Pradeep Ram and Sime-Ngando, 2010; Winter et al., 2010). The potential size and activity of the bacterial community and consequently the identity of the Virus-Host Systems (VHS) would also be driven by the bioavailability of nutritive resources (Sandaa et al., 2009). The bacterial stock and the number of VHS may depend upon the processes of species-dependent viriolysis and size-selective grazing (Bouvier and del Giogio, 2007; Church, 2008; Ovreas et al., 2003; Thingstad and Lignell, 1997; Winter et al., 2010).

Since imports of benthic organic matter and microorganisms have the potential to modify the dynamics of microbial entities and the functioning of the pelagic microbial food web, previous experiments have already tested the direct and indirect effects of marine sediment resuspension on planktonic bacteria and protozoans by using either intertidal sediment at two macrotidal sites [a marsh system (Hopkinson et al., 1998) and a muddy site (Garstecki et al., 2002)] or 20-m-deep sandy sediment (Wainright 1987; 1990). But viruses have always been excluded from these investigations. In the present study, we investigated the intertidal mudflats, which cover up to 60% of the macrotidal coastal area of the Marennes-Oleron Bay (France). At low tide, a complex and transient assemblage of both eukaryotic (as mainly epipelic microalgae, protozoans) and prokaryotic cells as well as viruses were associated on and in the surficial sediment in a mucilage matrix, i.e. the microphytobenthic biofilm (Guizien et al., 2013). A rising tide may more or less erode microbes living in this biofilm (Guizien et al., 2013). We designed experimental bioassays to investigate the potential alteration of the microbial interactions in the water column by the suspended biofilm. Using an erosion device (Orvain et al., 2007), the surficial substratum and the associated microorganisms living in the pore water or attached to sedimentary particles were eroded from a sediment core and added to pelagic water inocula. We then compared the temporal dynamics of microbes (virus, bacteria, HNF), bacterial and viral activity, fingerprint diversity and viral life strategies. The objectives were (1) to untangle the effects mediated by benthic organic matter and nutrients on microbial interactions from those of resuspended benthic microorganisms and (2) to examine whether there is a seasonal trend to the induced modification in food web functioning by targeting two seasons with contrasting productivity.

## 2 Materials and Methods

*2.1 Sediment and water sampling*

Water column samples were taken with the flat-bottom oceanographic barge "ESTRAN", at the sub-surface (1-m depth) in the Marennes-Oléron Bay, French Atlantic coast, at station E (7-10m depth range, Fig. 1). A few days before each experiment conducted on 3 March and 14 July 2008 (e.g., on 27 February 2008 and 9 July 2008, respectively), 50 L was collected at high tide, in order to produce < 30kDa ultrafiltered seawater by tangential flow through a 30-kDa polysulfone cartridge (Ultraslice support, Sartorius). This ultrafiltrate was conserved at 4 °C in autoclaved Erlenmeyer flasks until filling the flume of the erosion device and using it as pelagic water diluent in the experimental treatments (Fig. 2B). This ultrafiltrate served as diluent of pelagic inocula, either unenriched (in pelagic control treatments) or enriched with eroded sediment mimicking the erosion influxes (in erosion treatments). Sediment was sampled with an aircraft boat on 3 March and 14 July 2008 at low tide during the emersion of the Brouage intertidal mudflat located at the south end plume of the Charente estuary (Fig. 1). Cores of the 2-cm top surface sediment were collected using a 90-mm diameter manual coring Polyvinyl Carbonate device which was acid rinsed, ethanol disinfected and rinsed with field sediment before use. Cores were eroded within 30 minutes after collection. In parallel with the barge "ESTRAN", 25 L of seawater was sampled at mid-rising tide at station E and treated in the laboratory within 2 hours to obtain the different pelagic water inocula for the experiments (Fig. 2A).

2.2 Erodimetry

An erosion device was deployed near the beach of the Brouage mudflat (4 km) to enrich the ultrafiltered seawater in sedimentary particles, nutrients and biofilm. The erosion device developed by IFREMER and modified for this study is a straight recirculation flume named "Erodimeter" (Orvain et al., this issue). Briefly, two surficial sediment cores (90 mm in diameter) were settled into the eroding unit (a closed circuit which is connected to a pump). The sediment samples were directly transferred from cylindrical cores to the bottom of the flume of the erosion device. The < 30-kDa ultrafiltered seawater (free of microorganisms) was used to fill up the flume. Turbidity and fluorescence were recorded continuously to measure chl *a* and suspended particular matter to verify that initial erosion of microphytobenthic biofilm was well underway. In July, the flow discharge was incremented step-by-step with a succession of 20 steps, each one lasting 2 minutes. This procedure yielded bed friction velocities U* ranging from 0 to 6.6 cm.s$^{-1}$ with an increment of 0.34 cm.s$^{-1}$. In March, only 11 steps were applied to provoke an initial erosion phase (because of a lower critical threshold for biofilm erosion, u*$_{crit}$) and U* ranged from 0 to 4.1 cm.s$^{-1}$, with an increment of 0.35 cm.s$^{-1}$. The total volume of suspended sediment water (35 L of enriched ultrafiltrate) was taken from the eroding device at the end of the process and directly used for experiments.

*2.3 Inocula and diluents*

Water volumes were systematically prepared by sequential steps, i.e. a prior screening through 250, 25 and 10-µm nylon meshes followed by filtration through 1.2, 0.6 and 0.2-µm membranes (Sartorius, cellulose acetate) and eventually tangential ultrafiltration.

Four types of water served as diluents (Fig. 2B) after equilibration to the ambient seawater temperature: for control treatments the unenriched ultrafiltrate (P diluent) and for the erosion treatments the sediment suspension water (e.g., the P diluent enriched in the erodimeter and called -B diluent) which was size-fractionated by differential filtration through 10 µm (10µ-B diluent), 0.6 µm (0.6µ-B diluent) or ultrafiltered (9 L, benthic ultrafiltrate, -UFB diluent).

On the experiment day, two kinds of pelagic inocula were set aside during the filtration process of field water (Fig. 2A), 1.5 L each of < 10 µm water (10µ-I) and < 0.6 µm water (0.6µ-I) and maintained at *in situ* sea water temperature.

*2.4 Experimental design*

All treatments (Fig. 2C) consisted in 100 mL of pelagic inocula (0.6µ-I or 10µ-I; Fig. 2A) ten-fold diluted with one of the diluents (Fig. 2B) in 1-L Whirl-Packs. All treatments were placed in the dark to avoid the development of phototrophs in an incubation Shaking Cabinet (49 rpm, Certomat BS-1 Sartorius) for 116 h at 11.3 °C (March) and for 80 h at 19.9 °C (July). The agitation prevented settlement of resuspended matter. The experimental design differentiated two control treatments corresponding to pelagic assemblages (-P treatment) by using the P diluent to dilute either the 0.6-µ inoculum or the 10-µ inoculum. Two erosion treatments (-BP treatments) were prepared by mixing one inoculum with its corresponding size-class benthic diluent. To differentiate the pure effect of benthic microorganisms from mineral and organic matter impact (Low Molecular Weight compounds, LMW), inocula were diluted with the UFB diluent (-UFB treatments). Six treatments were thus tested, each in triplicate: 0.6-µ P, 10-µ P, 0.6-µ BP, 10-µ BP, and 0.6-µ UFB, 10-µ UFB. The 'HNF effect' (HNF factor) was evaluated for pelagic waters by comparing the 10-µ P treatment with the 0.6-µ P treatment that only contained heterotrophic bacteria and viruses. The total sampling period and the sub-sampling time was based on previous batch-culture experiments performed with water from the same environment (Auguet et al., 2009) to reach the plateau phase, which occurred faster in July (around 30 h post-inoculation) than in Marsh (> 90 h) and to qualify the lag phase which was three-fold longer in March (around 30 h) than in July (12 h). Sub-sampling was performed inside a Microbial Safety Cabinet to avoid contamination of the Whirl-Packs.

*2.5 Nutrient analyses*

Total Suspended Seston (TSS), Organic Matter and nutrients were measured according to Aminot (1983) in natural water sampled at an estuarine station and used to prepare inocula, in suspended sediment water during the erosion process and in treatment whirl-packs at $T_0$ and at the final point $T_{80}$ (in March) or $T_{116}$ (in July). Samples of 200 mL of seawater were filtered onto pre-weighted and pre-combusted (490 °C) GF/F (Whatman) filters; TSS corresponded to the increase in weight of filters after drying at 60 °C for 12 h and particulate mineral matter (PMM) to the weight of particles after a supplementary burning step of 2 h at 490 °C. Particulate organic matter (POM) was estimated as *POM = TSS – PMM*. Water samples for dissolved inorganic nutrient concentrations (nitrate [$NO_3$], nitrite [$NO_2$], phosphate [$PO_4$]) and dissolved organic carbon (DOC) were filtered through GF/F filters. Nutrient filtrates were stored at -20 °C until analysis using a TechniconIII (Bran+Luebbe) Auto-analyzer (Strickland and Parsons, 1972). After addition of conservatives (1/100 V, 1 N HCl), DOC filtrates were stored at 4 °C until analysis by high temperature catalytic oxidation (Cauwet, 1994) with a Shimadzu TOC 5000 analyzer. The percentage of labile C (%DOC) and the bacterial growth efficiency ($BGE_{DOC} = \Delta POC / \Delta DOC$) was estimated in 2006 by dilution culture of a < 0.8-µm inoculum in < 0.2-µm filtrate and using the CHN Erba Science 1500 Analyser for POC analysis. At the moment of the erosion process, wet sediment sub-samples were weighted and then dried at 60 °C for 12 h to assess the mass of the dry matter; POM was then estimated from the difference between the dry weight and the weight after a supplementary burn of 2 h at 490 °C.

*2.6 Microbial counts*

Subsamples for viral and bacterial counts were fixed with 0.02 µm-filtered formaldehyde (2% final concentration) at $T_0$, $T_{12}$, $T_{22}$, $T_{36}$, $T_{46}$, $T_{80}$ and $T_{116h}$ in March and at $T_0$, $T_6$, $T_{12}$, $T_{20}$, $T_{32}$, $T_{56}$ and $T_{80h}$ in July and stored for < 24 h at 4 °C to limit underestimation of counts linked to losses during the conservation of fixed samples (Wen et al., 2004). Samples were enumerated after 30-min SYBR Green I staining (Noble and Fuhrman, 1998) on a Zeiss Axioskop 2 Mot Plus microscope by screening 20 fields under blue excitation at 1000x magnification. To quantify small phytoplanktonic cells (< 10 µm) by cytometry, subsamples fixed in formaldehyde were flash-frozen and stored at -80 °C (Ory et al., 2011). Large phytoplanktonic cells preserved in alkaline lugol were counted by microscopy using Utermöhl settling chambers (Ory et al., 2011). For nanoflagellates, samples were fixed with paraformaldehyde (1% final concentration) and stored at 4 °C for < 1 wk. Cells were filtered on black polycarbonate 0.8-µm membranes (Nuclepore®), stained with DAPI (final concentration 2.5 mgL-1; Porter and Feig, 1980) and frozen at -20 °C until counting 50 fields under ultraviolet excitation.

*2.7 Lysogens induction*

Lysogenic tests were performed at $T_0$, $T_{22}$, $T_{36}$, $T_{46}$, $T_{80}$ and $T_{116\,h}$ in March and at $T_0$, $T_{12}$, $T_{32}$, $T_{56}$ and $T_{80h}$ in July. The frequency of lysogenic cells (FLC) was determined by induction of prophages, by adding mitomycine-C (0.5 µg mL$^{-1}$) to 5 mL of subsamples. Untreated samples served as controls. Samples were incubated at 11.3 °C (March) or 19.9 °C (July) for 12 h and then fixed with 0.02-µm-filtered formaldehyde (2% final concentration) before the viral and bacterial counting (see 2.6). FLC were estimated according to Bettarel et al. (2006) with a burst size of 33 (annual mean for the Charente estuary).

*2.8 Bacterial ectoenzyme activity*

Leucine aminopeptidase and β–glucosidase activities were measured using L-Leucine-7-amino-4-methylcoumarin hydrochloride (Leu-MCA, Sigma) and 4-methylumbelliferyl β-D-glucopyranoside (MUF-Glc, Sigma) as protein and carbohydrate model substrate, respectively, according to Hoppe (1993). A saturating concentration (1000 µM of Leu-MCA and 100 µM of MUF-Glc) was added to aliquot fractions of sampled water to determine the maximum enzymatic velocity ($V_{max}$) and the potential enzymatic activity per cell (specific $V_{max}$). Incubations were performed in the dark at *in situ* temperature for 7–14 h. Sodium dodecyl sulfate (1% final concentration) and $NH_4$-Glycin buffer (pH 10.5) were added to stop the respective reactions of aminopeptidase and β-glucosidase activity assays (Delmas and Garet, 1995) and to inhibit reactions, at the beginning of incubation, in negative controls. Samples were immediately frozen at -20 °C until fluorescence readings could be performed within a month (Kontron SFM 25 spectrofluorometer: excitation, 380 and 364 nm, emission 440 and 450 nm for Leu-MCA and MUF-Glc, respectively). Solutions of 7-amino-4-methylcoumarin (20 to 2000 nM, Sigma) and 4-methylumbelliferon (2 to 2000 nM, Sigma) were used as calibration standards.

*2.9 TTGE fingerprinting*

At $T_0$ and at the final point $T_{116}$ (March) and $T_{80}$ (July), bacteria (250 mL of an equal mix from the triplicated whirl-packs) were filtered on polycarbonate 0.2-µm membranes (47 mm, Whatman) after a pre-filtration on polycarbonate 3-µm membranes, if necessary, to remove suspended matter and large eukaryotic cells. Membranes were stored at -80 °C and the DNA was extracted by incubating (1) for 45 min at 37 °C in the presence of lysozyme (1 mg.ml$^{-1}$ final) and lysis buffer (Trisma-base 50 mmol.L$^{-1}$, Saccharose 7.5 mol.L$^{-1}$, EDTA 27 mmol.L$^{-1}$) (2) for 1 h at 56 °C under slow agitation in the presence of 20% SDS (1% final) and proteinase K (0.2 mg.ml$^{-1}$ final) and then purified using the 'DNeasy blood and tissue' kit (Qiagen). A fragment of 473 bp including the V6 to V8 regions of 16S rDNA was amplified using the primer set GC-968f and 1401r (Engelen et al., 1998) using 2 µL DNA, 1X Taq polymerase reaction buffer, 320 µM dNTP, 3 mM $MgCl_2$, 0.2 µM primers, 2U Taq polymerase (HotStart Qiagen) and the following PCR program: 94 °C for 15 min, 36 cycles of 90 °C for 30 s, 56 °C for 30 s, 72 °C for 30 s and a final extension step at 72 °C for 2 min (PTC-100, MJ Research Inc.). Amplicons were purified and concentrated on an Ultracel YM50 column and checked with a spectrophotometer (Nanodrop 2000, Thermo Scientific). Amplicons (400 ng) were loaded into an 8% (37.5:1) polyacrylamide (7 M urea) gel. Electrophoresis was run for 17 h at 68 V with 1.25x TAE (Tris-Acetate EDTA) buffer under a temperature gradient ranging from 66 to 69.7 °C incremented by 0.2 °C.h$^{-1}$. Gels were stained for 30 min in GelStar (Lonza, 0.5 µg.ml$^{-1}$ and 1.25x TAE) and bands (as operational taxonomic units, OTUs or species) were analysed using GelDoc and Diversity Database softwares (BioRad). The Shannon diversity index H' and the Pielou evenness index J' were calculated and the Dominance estimated as *D=1-J'*.

*2.10 RAPD-PCR fingerprinting*

Viral community structure was assessed exclusively at the final point $T_{116h}$ (March) or $T_{80h}$ (July) from samples combining the remaining triplicate incubation water. According to the current protocol used in our laboratory (Auguet et al., 2009; Ory et al., 2011), viruses were purified and concentrated to around 100 mL using 0.2-µm filtration and ultrafiltration. After storage at -20 °C, the viral concentrate was ultracentrifuged at 150,000g for 3.5 h at 4 °C (LE 70 Beckmann ultracentrifuge, rotor SW 28.1). The viral pellet was resuspended in SM buffer

[100 μl: 0.1 M NaCl, 8 mM MgSO$_4$ H$_2$O, 50 mM Tris-HCl, 0.005% glycerol (w/v)] and stored at 4 °C. Checks for nonviral DNA were performed by both DNAse testing and running a PCR with 16S rDNA primers. After virion decapsidation by heating at 60 °C for 10 min and rapidly cooling to 4 °C, random amplification was performed with the primer CRA 22 (MWG-Biotech Inc.). The PCR products were resolved and visualized according to Ory et al. (2011).

*2.11 Calculation of microbial productions*

Mean Viral Production (MVP) was estimated according to Luef et al. (2009) as the mean of all the net increases of each peak of viruses divided by the time of viral increase, over 116 h in March and 80 h in July. The same kind of estimation was applied for mean bacteria production (MBP) and production of heterotrophic nanoflagellates. The maximum yield (max-Y) of cellular production was estimated at the plateau as the net increase of bacterial cells with respect to T$_0$.

*2.12 Statistical analyses*

In order to access the significance of changes in microbial dynamics, two-way analysis of variance with Bonferroni post-tests (ANOVA) was carried out with PRISM 4 (Graphpad) software. Similarity between bacterial assemblages (or viral assemblages) was expressed as the similarity of Bray-Curtis considering the relative amount of amplified DNA associated with each OUT. The dendrogram was constructed using the Bray-Curtis similarity matrix and the group-average linking (Primer 6 software).

**3 Results**

*3.1 Pelagic food web structure*

A preliminary study was conducted in 2006 to evaluate the lability of DOC and the bacterial growth efficiency (BGE$_{DOC}$) in the outer estuary of the Charente River. In March 2006, bacteria took up 0.86 mg out of 4.85 mg.L$^{-1}$ DOC to produce 0.32 mg L$^{-1}$ of bacterial carbon and in July 1.89 mg out of 3.9 mg.L$^{-1}$ DOC to produce 0.49 mg.L$^{-1}$ C, indicating the presence of 18 and 49% of labile C (%DOC) with a BGE$_{DOC}$ of 37 and 26% in March and July, respectively. In March 2008, the DOC was 2.53 ± 0.08 mgC.L$^{-1}$, two-fold higher than in July 2008 (around 1.37 mgC.L$^{-1}$, Table S1). The N-nutrients concentration (mainly NO$_3$) was around two-fold higher in March than in July (73.65 ± 4.54 μmoles.L$^{-1}$ *vs.* 30.78 ± 4.54

µmoles.L$^{-1}$, Table S1) and phosphates were around 4-fold higher (0.87 ± 0.27 *vs.* 0.23 ± 0.12 µmoles.L$^{-1}$). The total number of phototrophic microorganisms in the natural water column was 9.73 x 10$^7$ small cells.L$^{-1}$ and 2.44 x 10$^4$ large cells.L$^{-1}$. In July, small cells were lower (2.79 x 10$^7$ cells.L$^{-1}$) and large cells higher (7.14 x 10$^4$ cells.L$^{-1}$). Viruses in March were 8.95 x 10$^6$ particles.mL$^{-1}$ and in July 1.70 x 10$^7$ mL$^{-1}$, while bacterial abundance in March was 1.36 x 10$^6$ cells.mL$^{-1}$ and in July 4.93 x 10$^6$ mL$^{-1}$ (Table S1). The *in situ* VBR was, thus, higher in March than in July (6.54 *vs.* 3.24). HNF were lower in March compared to July (3.36 x 10$^3$ *vs.* 8.29 x 10$^3$ cells.mL$^{-1}$).

*3.2 Suspended sediment water and water-treatments: initial conditions*

The critical thresholds for biofilm erosion were estimated to be 2.63 and 4.65 cm.s$^{-1}$ in March and July, respectively. In the field, maximum shear velocities were systematically obtained during the peak flood with an average of 3.3 cm.s$^{-1}$ in March and 4.7 cm.s$^{-1}$ in July (Orvain et al. this issue). In the seawater of the erosion device, the total particulate matter flow was 112 mg.m$^{-2}$.s$^{-1}$ in March and 23 mg.m$^{-2}$.s$^{-1}$ in July at the end of the experiment (after biofilm erosion). POM in the suspended sediment water was 32 mg.L$^{-1}$ in March and 15 mg.L$^{-1}$ in July, *vs.* 3 and 9 mg.L$^{-1}$, respectively, in the field water (Table S1, B *vs.* A). Inside the erosion device, the microphytobenthic biofilm was eroded into the suspended sediment water as indicated by counts of large microalgae (5.4 x 10$^4$ cells.L$^{-1}$ in March and 17.5 x 10$^4$ cells.L$^{-1}$ in July): their abundance appeared 2.3-fold higher than in the field pelagic water (data not shown). The resuspension of the biofilm was also translated into the occurrence of free bacteria, viruses and HNF (around 10$^5$ cells, 10$^6$ particles and 10$^2$ cells mL$^{-1}$ respectively, Table S1, *B whole*) and picoautotrophs (10$^2$ cells.mL$^{-1}$ in March and 10$^3$ cells.mL$^{-1}$ in July).

The dilution of the two types of pelagic inocula with their corresponding benthic diluents provided variations in the VBR (Table S1, C), in accordance with relative differences in the erodability of microorganisms and with their losses during the filtration steps (Table S1, A, B). The VBR in 0.6-µ treatments was always greater than in 10-µ treatments. Moreover, in March the VBR in the 0.6-µ PB treatment was greater than in the P treatment control (13.88 ± 1.88 *vs.* 8.32 ± 3.68, mean ± SD) but lower in July (8.73 ± 2.44 *vs.* 13.08 ± 3.70). VBR was always the lowest in the 10-µ PB treatments (3.28 ± 0.23 in March and 2.83 ± 0.97 in July). Within the nanoflagellate assemblage of the 10-µ BP treatment, the proportion of autotrophs was lower than in the pelagic assemblage of the 10-µ P and 10-µ UFB treatments (3.9 *vs.* 11.4 and 17.3% in March and 6.6 *vs.* 14.6 and 14.3% in July). Among HNF, the > 3-µm size class was more represented (36.0 *vs.* 24.5% in March and 39.3 *vs.* 31.9% in July). The initial inorganic nitrogen input (from the eroded sediment core) remained weak in the 0.6-µ UFB treatment compared to the control 0.6-µ P (Table S1, C), only +8.25% in March (from 72.64 to 78.64 µM) and +10.05% in July (from 32.25 to 35.49 µM). However, nitrites were more enriched than nitrates ($NO_2$: around +292% and +20% in March and in July, respectively; $NO_3$: around +5% and +13%, respectively; Table S1, C). The input in phosphate was negligible (around +2% and 0%, respectively). DOC-data were available only for experiments conducted in March (Table S1, C); the DOC in UFB treatments was nearly 10-fold higher than in the P treatments (up to 22-24 mgC.L$^{-1}$ compared to 2.5 mgC.L$^{-1}$) but also nearly two-fold higher than in the 0.6-µ PB treatment (14 mgC.L$^{-1}$). The DOC did not increase between P treatments and field water samples (Table S1, C *vs.* A).

*3.3 Batch-cultures*

The dynamics of bacteria, viruses and HNF over time (Fig. 3) were analyzed by comparing the Mean Bacterial Production (MBP), maximal net increase of bacteria (maxima yield, max-Y), Mean Viral Production (MVP) (Fig. 4) and Mean HNF Production.

3.3.1 Controls 0.6-µ P and 10-µ P

In absence of HNF (0.6-µ P), the MBP was 5.5-fold higher in March than in July (1.64 ± 0.3x10$^4$ *vs.* 8.98 ± 1.69x10$^4$ cells.mL$^{-1}$.h$^{-1}$; Fig. 4A, 4B). In March, the presence of HNF (10-µ P) significantly increased the MBP (ANOVA, $p < 0.01$) and tended to maintain the MVP (3.62 ± 1.19x10$^4$ *vs.* 4.01 ± 0.52 x 10$^4$ particles.mL$^-$1.h$^{-1}$; Fig. 4E, 4F) but increased the dominance within the viral species (Fig. 6). The max-Y significantly tripled between the two P treatments (ANOVA, $p < 0.01$) with a 28% decrease in DOC over 116 h in 10-µ P, instead of 1% in 0.6-µ P (Table S1, March C). In July, the bacterial growth was not modified by the presence of HNF (neither in terms of MBP nor max-Y; Fig. 3, Fig. 4). Only the MVP was somewhat stimulated (3.57 ± 1.31 *vs.* 5.64 ± 2.22 x 10$^4$ particles.mL$^{-1}$.h$^{-1}$; Fig. 4E, 4F) with concomitant changes in the dominance profile of viral species (data not shown). At the final time point in March and July, the abundances of HNF were similar (around 2 x 10$^3$ cells.mL$^{-1}$) in 10-µ P (Fig. 3) while autotrophic nanoflagellates had disappeared, inferring a mean production of 1.70 x 10$^3$ and 1.58 x 10$^3$ HNF cells.mL$^{-1}$.h$^{-1}$, respectively.

3.3.2 -UFB treatments *vs.* Controls: input of benthic < 30 kDa LMW compounds

In the absence of HNF, the MBP of pelagic assemblages in both seasons was stimulated by eroded LMW compounds (UFB treatments, Fig. 4A, 4B). The increase was only significant in

March (ANOVA, $p < 0.001$, Table 1) and was corroborated by a 40% decrease in DOC (Table S1, C). However, Max-Y significantly increased (ANOVA, $p < 0.001$, Table 1) in both seasons. MVP tended to increase only in March (Fig. 4E). Thus, at $T_{116}$, the viral abundance was 6-fold higher ($6.63 \pm 1.33\ 10^6$ particles.mL$^{-1}$) than at $T_0$ ($1.09 \pm 0.29\ 10^6$ particles.mL$^{-1}$), while in the presence of HNF the increase leveled out at $1.39 \pm 0.28\ 10^6$ particles.mL$^{-1}$ (Fig. 3). In July in the presence of pelagic HNF, the MVP tended to double (10-µ UFB *vs.* 10-µ P, Fig. 4F).

Even if in March in the presence of HNF, the benthic compounds alone increased the MBP from $5.8 \pm 0.5$ to $11.4 \pm 0.8$ x $10^4$ cells.mL$^{-1}$.h$^{-1}$ (10-µ UFB *vs.* 10-µ P; ANOVA, $p < 0.001$, Table 1), concomitant with a decline in DOC (-33% *vs.* -28%, Table S1, C), the max-Y changed little (x1.2-fold, Fig. 4C) and the dynamics of HNF did not change (Fig. 3). However in July, the 2.6-fold enhancement of MBP (ANOVA, $p < 0.05$; Table1) was accompanied by a significant 2.4-fold increase in max-Y (ANOVA, $p < 0.01$; Fig. 4D, Table 1), and by a lesser growth of HNF than in the 10-µ P treatment (Fig. 3). HNF were $1.0$ x $10^3$ cells.mL$^{-1}$ at $T_{80}$. Compared with the 0.6-µ UFB treatment, the HNF inferred a significant negative impact on the max-Y of bacteria (ANOVA, $p < 0.01$).

The LMWC input increased the peak of the aminopeptidase activity in March at $T_{46}$ and in July at $T_{12}$, concomitantly to the end of the lag phase for bacterial growth (Fig. 3) and more markedly in the absence of HNF (Fig. S1): 2180 and 1100 amol.cell$^{-1}$.h$^{-1}$ (in 0.6-µ UFB and 10-µ UFB treatments, respectively) *vs.* 600–800 amol.cell$^{-1}$.h$^{-1}$ in the Controls. The β-glucosidase activity remained unchanged in March while in July, mainly in the presence of HNF, a peak occurred at $T_{12}$ (from 0.5–2 amol.cell$^{-1}$.h$^{-1}$ at $T_0$ to 7 *vs.* 5 amol.cell$^{-1}$.h$^{-1}$ in the 10-µ UFB *vs.* 10-µ P Control).

The structure of the bacterial community was modified by the input of eroded LMWC as shown by the clustering of the TTGE fingerprints and the diversity indices (Fig. 5). In March at the final time point $T_{116}$, the Shannon-Weaver index increased while the dominance decreased. In July, the dominance profile was more marked with a concomitant disappearance of OTUs, mainly in the presence of HNF (Fig. 5).

3.3.3 -BP treatments *vs.* other treatments: input of microphytobenthic biofilm

In both seasons, the benthic microorganisms tended to enhance the MBP in -BP treatments (Figs. 4A, B) as compared to their respective -P treatment control but less than expected by the increase in benthic matter, as compared to their respective UFB treatment. This increase was only significant in March in the absence of HNF (ANOVA, $p < 0.01$; Table1).

In 0.6-µ BP treatments in both seasons, the increase in max-Y (Figs. 4C, D) was limited by the presence of benthic microorganisms, but only significant in July (ANOVA, $p < 0.01$; Table1). Moreover in March, the presence of the bigger benthic microorganisms, (0.6–10 µm in size), restrained max-Y to values below those of the 10-µ P treatment control, (Figs. 4C, D). Taking into account all six treatments, the variation in max-Y of bacteria depended on both the type of treatment and the presence/absence of HNF. In March, the HNF factor alone was not significant, contrary to July (ANOVA, $p < 0.01$, Table1). In fact, both the treatment and HNF factors interacted. In March, this interaction accounted for 27.1% of the total variance and in July for 9.7% (ANOVA, p = 0.005 and 0.010, respectively; Table1). No significant interaction was detected for MBP and MVP, neither in March nor in July (Table1). In March, as in the 10-µ P treatment, HNF began to increase at $T_{46}$, yet at the final time point $T_{116}$ they were five-to-ten-fold more abundant than in the 10-µ P and 10-µ UFB treatments (18 x $10^3$ vs. 2–4 x $10^3$ cells.mL$^{-1}$; Fig. 3). In July, their growth began at $T_{32}$, but only at $T_{56}$ in 10-µ P. The difference in mean production was more marked in March (1.8 x $10^4$ vs. 1.7 x $10^3$ cells.mL$^{-1}$.h$^{-1}$) than in July (3.7 x $10^3$ vs. 1.8 x $10^3$ cells.mL$^{-1}$.h$^{-1}$). The HNF growth was mainly due to large cells in the BP treatments (cells > 3 µm: 79% in March at $T_{116}$ and 73% in July at $T_{80}$), while in the P treatments 89–100% of the HNF were small (< 3 µm) cells. In the 0.6–10-µm size range, picoeukaryotes and cyanobacteria were also originally present, and declined more rapidly in the presence of benthic microorganisms (loss of around 80–92% in 10-µ BP vs. 42–51% in 10-µ P and 10-µ UFB treatments).

The MVP was enhanced by biofilm in March, but only significantly in the absence of HNF (ANOVA, $p < 0.05$; Table1). In July, the MVP tended to be intensified only in the presence of HNF and the first peak of viruses occurred earlier, at $T_6$ (Fig. 3). During the first 12 h of incubation (Fig. 3), the net viral production in July was positive only in the 10-µ P and 10-µ UFB treatments (1.06 ± 0.86 $10^6$ and 1.30 ± 0.50 $10^6$ particles.mL$^{-1}$, respectively). In March, viruses peaked first at $T_{12}$ in the 0.6-µ P and 10-µ BP treatments, but the net viral production over 12 h was higher in the presence of biofilm (6.91 ± 4.71 $10^5$ and 1.50 ± 0.06 $10^6$ particles mL$^{-1}$, respectively). In March, the inducible lysogens appeared later in the -P treatments (10-µ-P: $T_{46}$, 1.81% and 0.6-µ-P: $T_{116}$, 0.32%) than in the -BP treatments, (10-µ-BP: $T_0$, 0.54% and 0.6-µ-BP: $T_{22}$, 3.88%). In July, the presence of biofilm induced the lysogenic decision from $T_0$ on, irrespective of HNF (6.81% in 0.6-µ BP and 2.35% in 10-µ BP treatments). In the P treatments, inducible lysogens were only detected at $T_{12}$, when the VBR declined due to the bacterial growth phase (1 and 0.79% in the 0.6-µ P and 10-µ P treatments, respectively, Table S2).

In March, the presence of benthic microorganisms in the -BP treatments reduced the lag phase of the bacterial growth, and the aminopeptidase activity peak was inhibited (merely 80 amol.cell$^{-1}$.h$^{-1}$), while it occurred in the four other treatments at $T_{46}$, with the onset of bacterial growth (300–400 amol.cell$^{-1}$.h$^{-1}$; Fig.S1). However in presence of HNF, the enzymatic activity began increasing later, from $T_{80}$ on, to attain 680 amol.cell$^{-1}$.h$^{-1}$ at $T_{116}$, concomitant to the important increase of the β-glucosidase activity (from 0.5 at $T_{80}$ to 4 amol.cell$^{-1}$.h$^{-1}$ at $T_{116}$). In July (Fig. S1), the aminopeptidase activity leveled out at 600 amol.cell$^{-1}$.h$^{-1}$ during the instantaneous bacterial growth phase in the 10-µ BP and 0.6-µ P treatments, while in the other treatments the activity increased until the end of the bacterial lag phase (Fig. 3).

When comparing the initial ($T_0$) bacterial community structure of -P treatments with -BP treatments in March, the Shannon-Weaver index increased and dominance declined in the latter (Fig. 5). By contrast, in July the number of OTUs declined while dominance increased. In both seasons, the clustering of the end-assemblage structure distinguished the 10-µ BP assemblage, characterized in March by the highest dominance and in July by the lowest dominance of all treatments (Fig. 5).

The structure of the viral communities evolved to form the same four clusters at the final time points $T_{116}$ in March and $T_{80}$ in July (Fig. 6). At the threshold of 45% of similarity, the 0.6-µ P and 0.6-µ BP assemblages formed two singletons and the two 10-µ BP and 10-µ UFB assemblages were distinguished from the cluster encompassing the 0.6-µ UFB and 10-µ P assemblages. The number of species tended to be reduced in the 0.6-µ BP assemblage and to be enhanced in 10-µ BP assemblage as compared to their respective control. The dominance was highest in March in the 0.6-µ BP and in July in the 10-µ BP treatment.

## 4 Discussion
### 4.1 Methodological considerations

Our experimental design allowed the untangling of the direct and indirect effects of benthic organic and inorganic matter on trophic interaction from those of resuspended benthic microorganisms. The experiments were also designed to compare the effect of benthic biofilm with that of HNF on the microbial interactions. The results validate the choice of MBP as a good descriptor of bacterial productivity, a proxy of growth rate alterations. However, it does not clearly indicate the trade-off between bottom-up (resources) and top-down (grazers, virus) controls, which is best represented by the maximum yield of cell biomass (max-Y, i.e., the maximal net increase). Moreover, while the analysis of diversity by fingerprinting only allows access to the dominant populations of microbes (and one fingerprinting band may contain more than one OTU), the comparison of fingerprints is relevant to monitor changes in the assemblage structure (Auguet et al., 2009; Hewson and Fuhrman, 2006; Ory et al., 2011).

While the results confirmed the resuspension of benthic microorganisms and POM in the erodimeter water, the resuspension of LMW compounds (DOC, nutrients) and their impact on food web functioning may be doubtful. Nitrites enriched the erodimeter water more than the other inorganic nutrients. Likewise, the DOC of benthic origin increased initially in the BP and UFB treatments, as compared to the P treatments. The unexpected higher level of DOC in the UFB treatments may be due to a combination of two factors: directly from mucilage of microphytobenthic biofilm and indirectly by the disruption of the eroded mucilage or cells (POM) during the filtration steps. Thus, our experimental design may have accentuated the physical transformation of POM to DOM, biasing the original POM/DOM ratio present in the suspended sediment water. Nevertheless, in seawater, the POM-DOM equilibrium has been shown to be instable with physical and biological transformations between the two stages (Verdugo et al., 2004).

*4.2 Environmental context and trophic links in the water column*

The initial conditions of batch-cultures mirrored seasonal differences in the erosion fluxes of the microphytobenthic biofilm and in the functioning of the benthic and pelagic food webs. Ecological network analysis and linear inverse modeling showed that in March an excess of microphytobenthic production and the high carbon recycling by bacteria would generate a pool of carbon and nutrients potentially exportable to the water column at high tide. By contrast in July, a larger proportion of the photosynthetic carbon (67, *vs.* 25% in March) would be integrated into the benthic food web (Saint-Béat, 2012).

In the water column of the Charente estuary, during the periods of the multivorous food web of March and July (Ory et al., 2010), the DOC peaked concomitantly with the phytoplankton growth, notably in March, but the availability of inorganic nutrients was always lower in July (Montanié, personal data). Based on earlier observations (2006), we can postulate that in July the bacterial growth may rely less on grazing products than in March because of a more labile DOM pool despite its lower concentration. Thus in 2008, HNF and the grazer-mediated resource (regenerated matter) enhanced the bacterial growth in March but not in July. By acting on the bacterial activities and diversity, the availability of inorganic and organic matter determines the identity of the virus-host systems (evenness, J'; dominance 1-J'). The HNF bacterivory and viral lysis regulate the number of the virus-host systems (specific richness, H') either synergistically or antagonistically (Ory et al., 2010; Ovreas et al.,

2003; Sandaa et al., 2009). In return, the regenerated DOM (viral lysis or protist grazing products; Middelboe and Jørgensen, 2006; Nagata and Kirchman, 1996) could interfere in maintaining the bacterial growth and activity, thereby, supporting viral production (Pradeep Ram and Sime-Ngando, 2008; Šimek et al., 2001; Weinbauer et al., 2007; Weinbauer et al., 2003; Winter et al., 2010).

In March 2008, the presence of HNF led to a great dominance among viral species together with the prevalence of lysogenic decision and a more diversified bacterial community (higher H'). In July, HNF enhanced the viral production and dominance and the two predators reduced the bacteria and their diversity (lower H') without changing their evenness (equal J'). Thus in the pelagic system, the export of matter to higher trophic levels potentially differs between March and July. In March, the carbon may be channeled out by HNF, while in July it would be more retained in the DOM-Bacteria-Virus loop (Fig. 7) in concordance with higher bacterial growth efficiency in March and greater DOM remineralization in July.

*4.3 Does microphytobenthic biofilm modify the microbial interactions and diversity?*
*4.3.1 Benthic LMW-Compounds*

Benthic sediment would be a source of labile DOM and inorganic nutrients, important for supporting bacterioplankton production (Hopkinson et al., 1998; Wainright, 1987) and indirectly for HNF (Garsteki et al., 2002). In March, the responses of bacteria to the input of benthic compounds confirmed the relative limitation in labile-DOC that may be leveled down independently by regenerated products or by benthic matter. While in periods of low availability in DOC and phosphates (July), the structuring effect of the nutritive resources on bacterial growth exceeded those of the HNF-predation. In nutrient-depleted areas, the bacterial communities would be mainly structured through competition for nutrients in accordance to the "killing the winner" (KtW) concept (Winter et al., 2010). Nevertheless, in both seasons, HNF and benthic compounds acted antagonistically, limiting the beneficial effect of nutrients on the maximal yield of bacteria and the HNF growth.

We reported, in the enriched waters, two contrasting interactions corroborating the hypothesized increase in the resistance of bacteria to HNF predation in DOC-rich waters (Thelaus et al., 2008). (1) In the absence of suspended matter, resistance occurred in March and the regenerated grazing products were sufficient for the growth of grazing-resistant bacteria and, thus, the benthic compounds did not further enhance bacterial yield nor HNF growth (2) In July, the HNF-resistant bacteria benefited from benthic compounds, reducing the availability of bacterial prey for HNF. The suspension of benthic compounds did not change alone the bacterial production but significantly increased max-Y, inferring selection and growth of K-strategist bacteria already adapted to this nutrient supply and able to rapidly catabolize carbohydrates or others biopolymers liberated from microphytobenthic mucilage. In fact, sediment resuspension changes the quality of POM, i.e., the labile versus refractory fractions and the protein/carbohydrate ratio, which may interfere with the catabolism of bacteria (Pusceddu et al., 2005).

Direct as well as indirect links characterized the interactions between bacteria and the predator guild of bacteria (virus and HNF). Without HNF, benthic compounds tend to stimulate the viral production in March, congruent with the KtW prediction of the viral multiplication at the expense of the most active bacteria (Winter et al., 2010). In the presence of the two predators, the greater the positive HNF effect was on viruses (Bonilla-Findji et al., 2009; Zhang et al., 2007), the more the mean viral production was stimulated by the presence of benthic compounds (July *vs.* March). In fact, both virulent and temperate phages may coexist in the pelagic inocula, and viruses adapted their lifestyle to variable nutrient concentrations and VBR (Pradeep-Ram and Sime-Ngando, 2010). In July, benthic resources and predation may together drive changes in bacterial phenotypes and in interactions mediated by physiological traits affecting the viral multiplication (Miki and Jacquet, 2008). At $T_{12}$ under predation pressure, the viral lifestyle decision shifted to lysogeny when the first peak of viruses was attained and the growth of resistant bacteria began. But the presence of suspended resources resulted in an anticipated lysogenic decision (at $T_0$). The competition between virions and nutrients for the site of adsorption on bacteria may govern this choice and intensify the initial selection of the more adapted bacteria to this new availability and quality of nutrients. Then a trade-off would occur between the lytic multiplication of the viruses and the growth of their host (Winter et al., 2010). In consequence, in July in the presence of HNF and benthic compounds, the bacterial community composition became less diversified and the structure was less even.

By contrast in March, the lysogenic conversion depended rather on HNF predation. HNF and benthic resources acted antagonistically on the viral dynamic, through the predation of the bacterial host and a potential enhancement of the intraguild predation of HNF on infected bacteria (Miki and Jacquet, 2008). However, the bacterial community composition became more diversified and the structure was more even. Benthic resources may favor the emergence of cryptic species among the bacterial community, resistant to grazers and permissive to viral infection. In return, viruses contributed to maintain the bacterial diversity (evenness and richness) in accordance with the KtW prediction.

*4.3.2 Bulk Biofilm*

Biofilm microorganisms (< 10 µm in size) enhanced the effects of benthic matter, favoring mainly in July a higher viral production and in March HNF growth. However, the maximum cell biomass of bacteria was strongly limited compared to those with benthic matter alone (Fig. 4). This limitation may ensue from the biased differences in initial DOC due to filtration and from changes in the quality of hydrolysable POM. Even so in the absence of HNF, the input of bulk microphytobenthic biofilm increased the mean production of bacteria and the maximal yield. In March, lytic production was significantly enhanced due to a sufficiently high contact rate, notwithstanding a strong episode of lysogeny at $T_{22}$. Moreover in both seasons in the presence of HNF, the lysogenic life-style coexisted with lytic production from $T_0$ on. If we cannot exclude the importation of benthic lysogens, the more plausible explanations to the immediate occurrence of inducible lysogens were changes in the VBR and in the level of competition for nutrients during the exponential growth of bacteria. Like in oligomesotrophic Lake Pavin, the lysogenic decision related rather to nutrient availability in the poorer nutrient environment of July (Pradeep-Ram and Sime-Ngando, 2010). However, benthic bacteria may also outcompete the pelagic bacteria since (1) the large fast-growing bacteria of benthic origin, as already shown by Wainright (1987), may be nitrifying species able to oxidize the benthic nitrites to nitrates (Fig. S2), (2) in March irrespective of HNF, the bentho-pelagic assemblage developed without an increase of aminopeptidase activity during the exponential growth phase (Fig. S1), and then the nutrition of the bacterial community shifted towards the end of batch-culture from labile DOC to hydrolysed polymers, and (3) in July without HNF, the pelagic assemblage of bacteria would be relatively adapted to using the ambient labile DOC. By changing the bioavailability of organic matter [the bioavailable fraction of the protein and carbohydrate pools (Pusceddu et al., 2005)], the input of benthic matter and benthic bacteria delayed the bacterial growth phase until the bacterial assemblage produced enough ectoenzymes to hydrolyze the new benthic polymers. Moreover, the energetic cost of extracellular hydrolysis (Del Giorgio and Cole, 1998) and the competition for resource may also explain the observed delay.

It is also noteworthy that the pico-phytoplankters, which were surviving in the dark, declined more in the presence of benthic grazers. The HNF grazing on picoeukaryotes and cyanobacteria may limit their diet upon bacteria and, thus, reduce the matter flow within the microbial loop (Van Wambeke, 1994).

Bacterial dynamics and microbial interactions within the benthic-pelagic assemblages were congruent in July with the strong influence of availability of organic and inorganic matter and in March with the influence of both HNF and matter. In both seasons, HNF influenced the temporal changes in the benthic-pelagic assemblage. Grazers may, thus, modify the resource control of bacterial diversity (Bonilla-Findji et al., 2009). In March in their absence, the assemblage evolved to a higher and even diversity, while their very high growth selected for less species of bacteria with high dominance despite a more even distribution of viruses. However in July, the decrease in specific richness of bacteria was not dependent on the presence/absence of HNF even if the presence of HNF decreased the dominance in line with a less diversified viral community. Thus, within the benthic-pelagic assemblage, viruses, bacteria and HNF may evolve relatively more in phase in March than in July, where viral lysis dominated bacterivory impacts.

**5 Conclusions:**

To our knowledge, we report the first experimental conclusions about the impact of microphytobenthic biofilm suspension on the interactions between viruses, bacteria and HNF in the water column. Bacterial and viral structures were modified by benthic biofilm suspension attesting to significant changes in the active virus-host systems and adaptations of viral lifestyle to environmental fluctuations. In March, the regulation of the bacterial community was complex as a result of competition for nutrients and of the interactions with HNF and viruses. Benthic biofilm suspension unevenly reduced the specific diversity. In July, the competition for resources governed the dynamics and the diversity of bacteria. Input of benthic biofilm selected for certain species and thus declined evenly the species richness.

*In situ* in March, viruses, bacteria and HNF along with organic and inorganic matter may be eroded more frequently from the mudflat than in July (Orvain, pers. comm.). Our results indicate that biofilm suspension lubricates the pelagic microbial web functioning, tending to export matter in March and to retain it in July. During nutrient abundance in March, biofilm input tends to increase the carbon flow through the microbial loop and, thus, facilitates the export towards the classical food web via the predators of HNF (Fig. 7). By contrast during nutrient depletion (July), biofilm input tends to stimulate the viral loop and reduces, in synergy with a strong grazing pressure, the size of the bacterial compartment, maintaining the carbon flow within the microbial loop (Fig. 7).

Nevertheless, to complete the picture, it would be necessary to consider the phototrophic components, the bioavailability of POM and DOM as well as the grazers of HNF, characterizing their reactivity to biofilm suspension and their interaction within and with the pelagic microbial food web.

**Acknowledgements**

This study was supported by the French ANR (National Research Agency) through the VASIREMI project "Trophic significance of microbial biofilms in tidal flats" (contract ANR-420 06-BLAN-0393-01). P. Ory's work was supported by a joint Ph.D. fellowship from the CNRS and the Région Poitou-Charentes. The authors thank M. Bréret (La Rochelle, LIENSs, UMR 7266), Laetitia Birée (Caen, CNRS, FRE3484 BioMEA, CNRS-INEE) and Jocelyne Caparros (Banyuls, LOMIC, UMR7621, CNRS/UPMC) for DOC and nutrient analyses and P. Catala (Banyuls for Cytometry). We appreciate the valuable comments and suggestions from three reviewers.

# Legends

Fig. 1: Study area in the Marennes-Oleron Bay (French Atlantic coast, France) characterized by a 2–5 m tidal range. Water sampling was performed at Station E, located in the outer part of the Charente River estuary (7–10 m depth). Sediment coring was conducted on the Brouage mudflat during emersion at 2 km from the beach.

Fig. 2: Flow chart of the experimental design. A: Preparation of inocula from pelagic water sampled at Station E. B: Preparation of diluent water from pelagic water (P-diluent), and of B-diluents from suspended sediment water (10µ, 0.6µ and UFB diluents). C: preparation of treatment whirl-packs as two purely pelagic controls (0.6-µ P and 10-µ P), two benthic-pelagic treatments (0.6-µ BP and 10-µ BP) and two pelagic treatments enriched with benthic matter (0.6-µ UFB and 10-µ UFB).   For details see the main text.

Fig. 3: Microbial Kinetics. Two pelagic inocula (< 0.6 µm and < 10 µm water) were diluted either with pelagic ultrafiltrate (0.6-µ P, 10-µ P treatments) or with B diluents, i.e., pelagic ultrafiltrate enriched with benthic biofilm and size-fractionated (0.6-µ BP, 10-µ BP treatments) or ultrafiltered (0.6-µ UFB, 10-µ UFB treatments).  Abundance of bacteria and viruses in March (left) and July 2008 (right). Abundance of heterotrophic nanoflagellates (HNF) in < 10-µ treatments.  (Mean ± SD, n = 3).

Fig. 4: Parameters of the kinetics of bacteria and viruses in March (left) and July 2008 (right). Six treatments were assembled: two purely pelagic controls (0.6-µ P and 10-µP), two benthic-pelagic treatments (0.6-µ BP and 10-µ BP) and two pelagic treatments enriched with benthic matter (benthic ultrafiltrate, 0.6-µ UFB and 10-µ UFB). For bacteria: Mean bacterial production MBP = mean of all the net increases of the bacterial peaks divided by the time elapsed, over the duration of the incubation (116 h in March and 80 h in July); Max-Y = the maximum yield of cell biomass at the plateau (= net increase of cells). For Virus: Mean viral production MVP (as for bacteria). Mean of triplicates ± SE.

Fig. 5: Clustering of bacterial assemblage structure at T0 and at the final time point T116 (March 2008) or T80  (July 2008) for the six treatments. Dendrograms using group-average linking of Bray-Curtis similarities. Control-samples at the final time point and 10-µ BP sample are shaded similarly in boxes. Data from the TTGE fingerprints of amplicons of the V6-V8 variable regions of 16S RDNA. H' = Shannon-Weaver diversity index; J' = Pielou evenness index.

Fig. 6: Clustering of the viral community structures at the final time point T116 (March, 2008) and T80 (July, 2008) for the six treatments. Dendrogram of Bray-Curtis similarities. Samples were shaded in four greyed boxes of similarity. Data from PCR-RAPD electrophoresis. H' = Shannon-Weaver diversity index; J' = Pielou evenness index.

Fig. 7: Conceptual schema of the benthic-pelagic fluxes within the microbial food web. Dashed lines correspond to matter fluxes through the microbial loop (ML) and full lines to the viral shunt of matter (VS). The variations in line and box sizes express the consequences of microphytobenthic biofilm input, materialized in black compared to the initial gray situations. In March (left), the biofilm tends to increase the carbon flow through the microbial loop and, thus, facilitates the export towards the classical food web via the predators of HNF. In July (right), the biofilm tends to stimulate the viral loop and reduces, in synergy with a stronger grazing pressure, the size of the bacterial compartment, maintaining the carbon flow within the microbial loop.

Table 1: Significance of two-factors Model I ANOVA on the mean bacterial production (MBP), mean viral production (MVP) and the maximal net increase of bacteria (maxY, maximum yield at the plateau) for experimental data of March and July 2008. Nanoflagellate (HNF) and treatment factors and their interactions were tested with the F statistic. *A posteriori* multiple comparisons were made by a Bonferroni test with three treatment levels: controls (0.6-µ P and 10-µ P treatments) *vs.* benthic low molecular weight compounds LMWC (0.6-µ UFB and 10-µ UFB treatments) and resuspended biofilm (0.6-µ BP and 10-µ BP treatments). *p* significance: * < 0.05, ** < 0.01, *** < 0.001, **** < 0.0001.

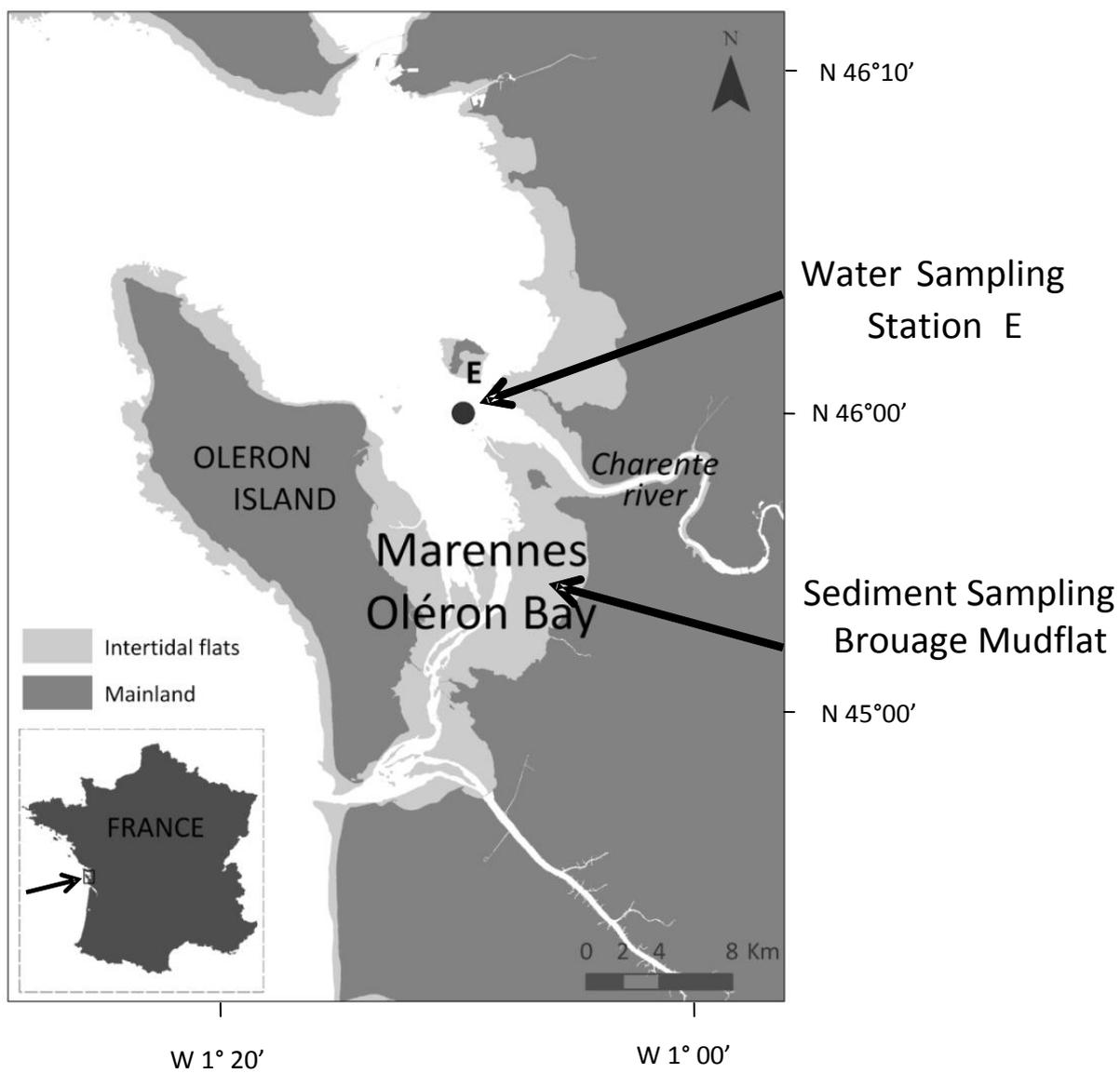

**Fig. 1**

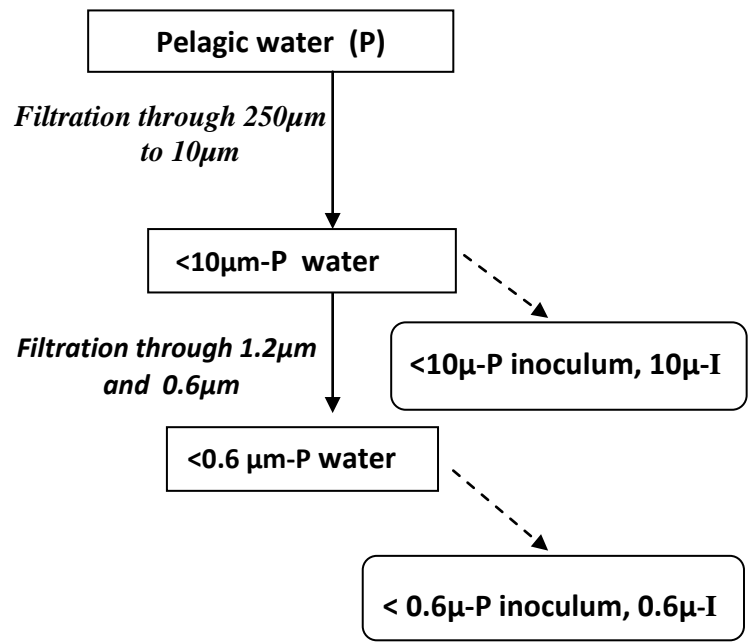
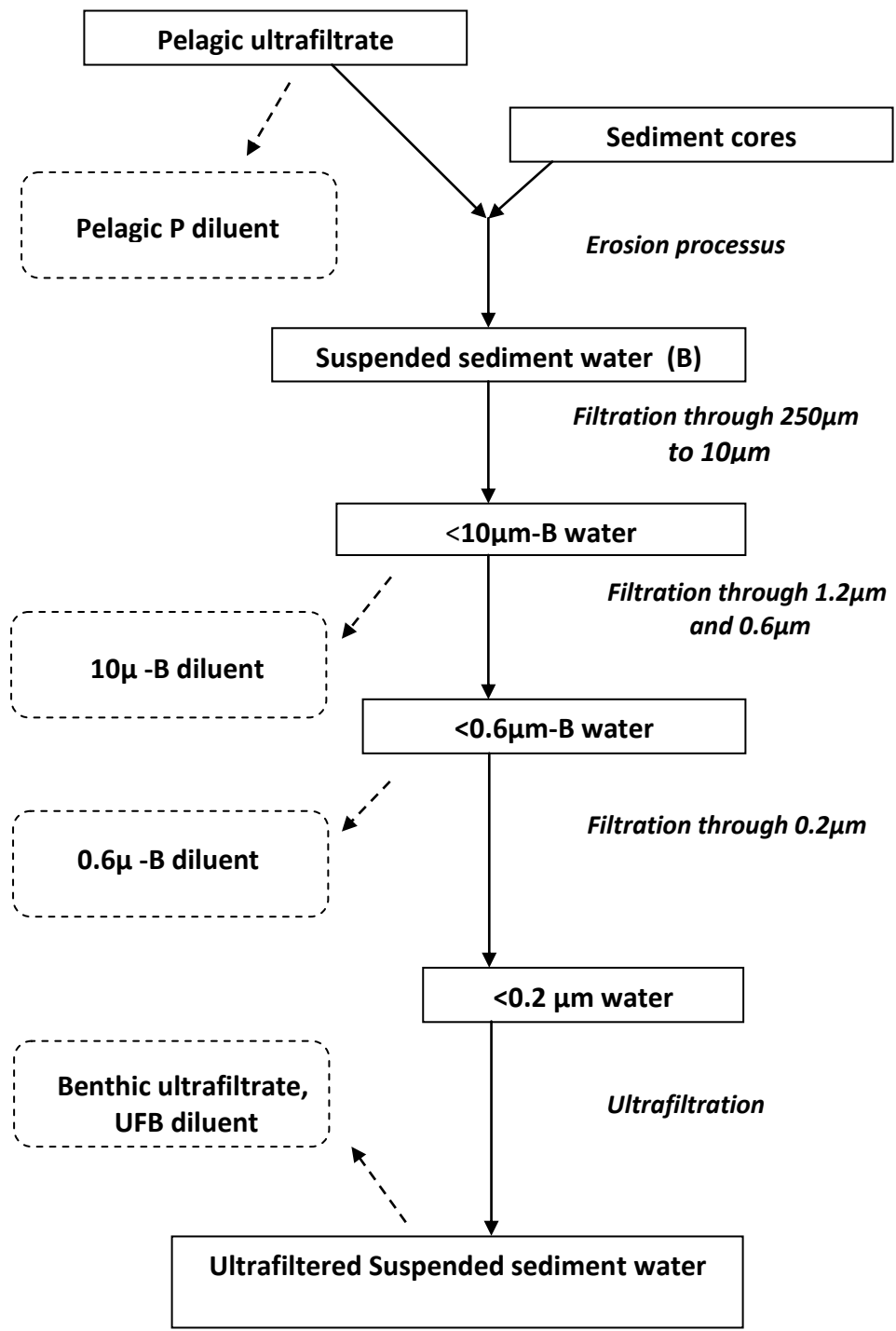
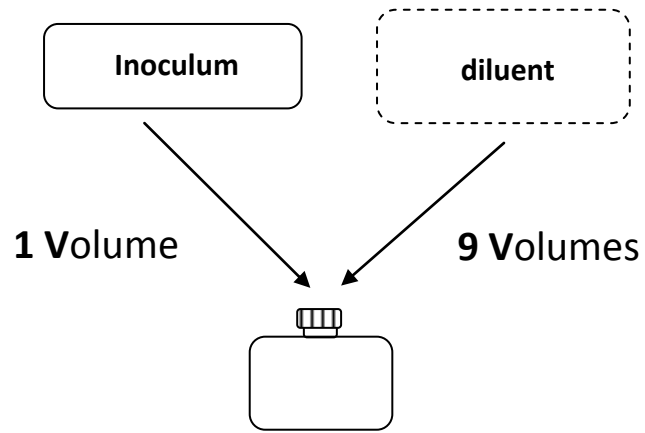

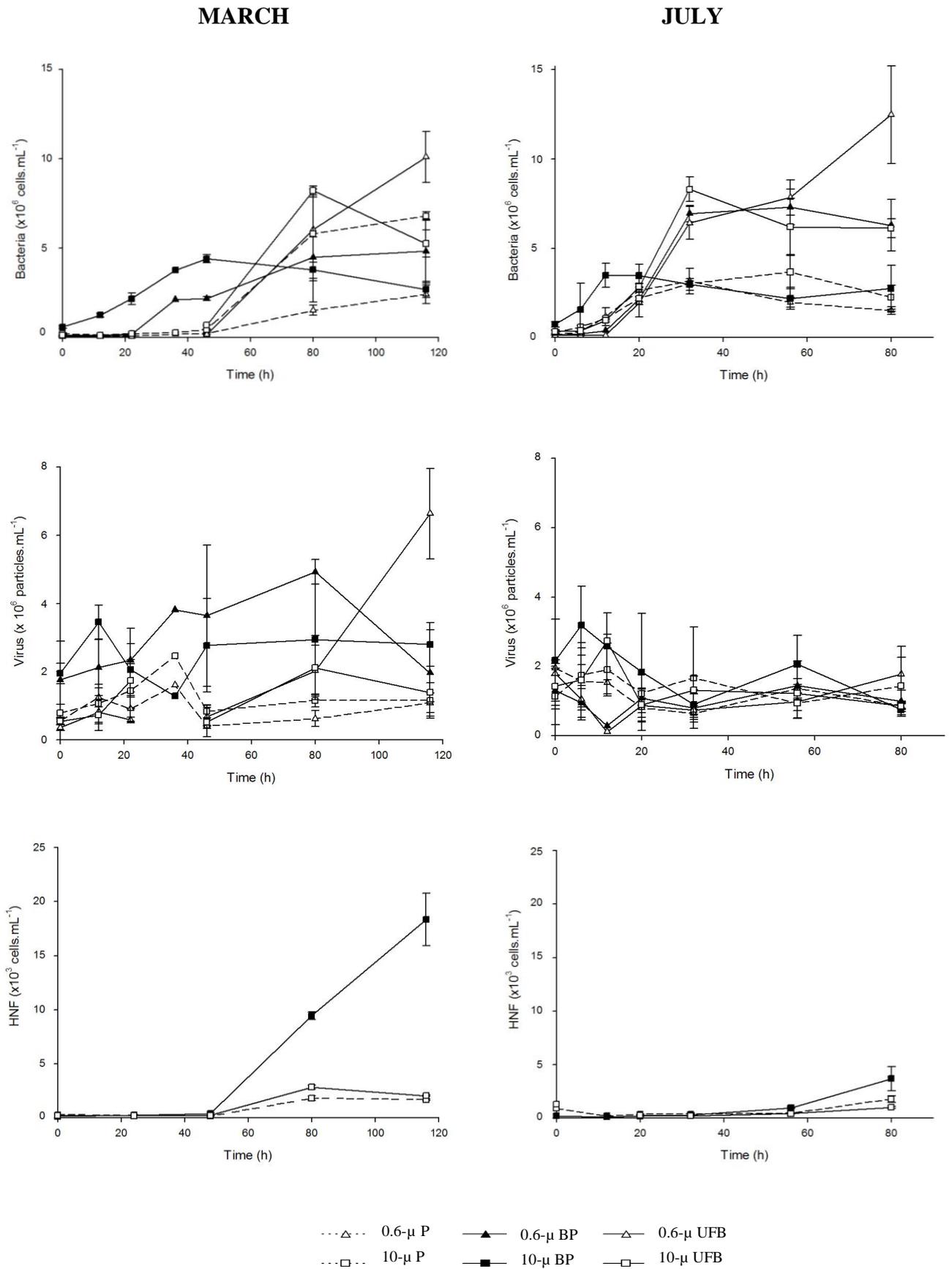

**Fig. 3**

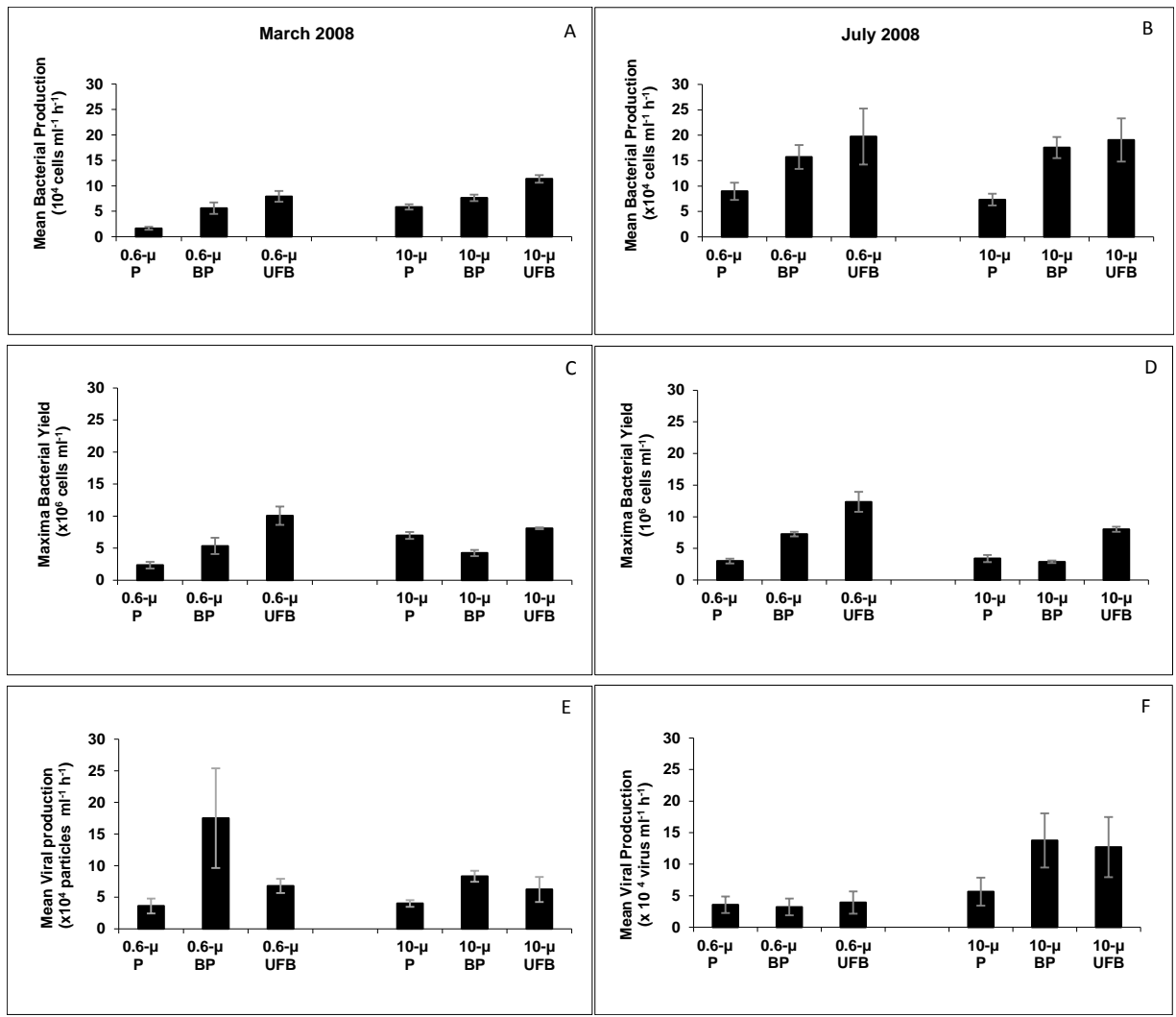

Fig. 4

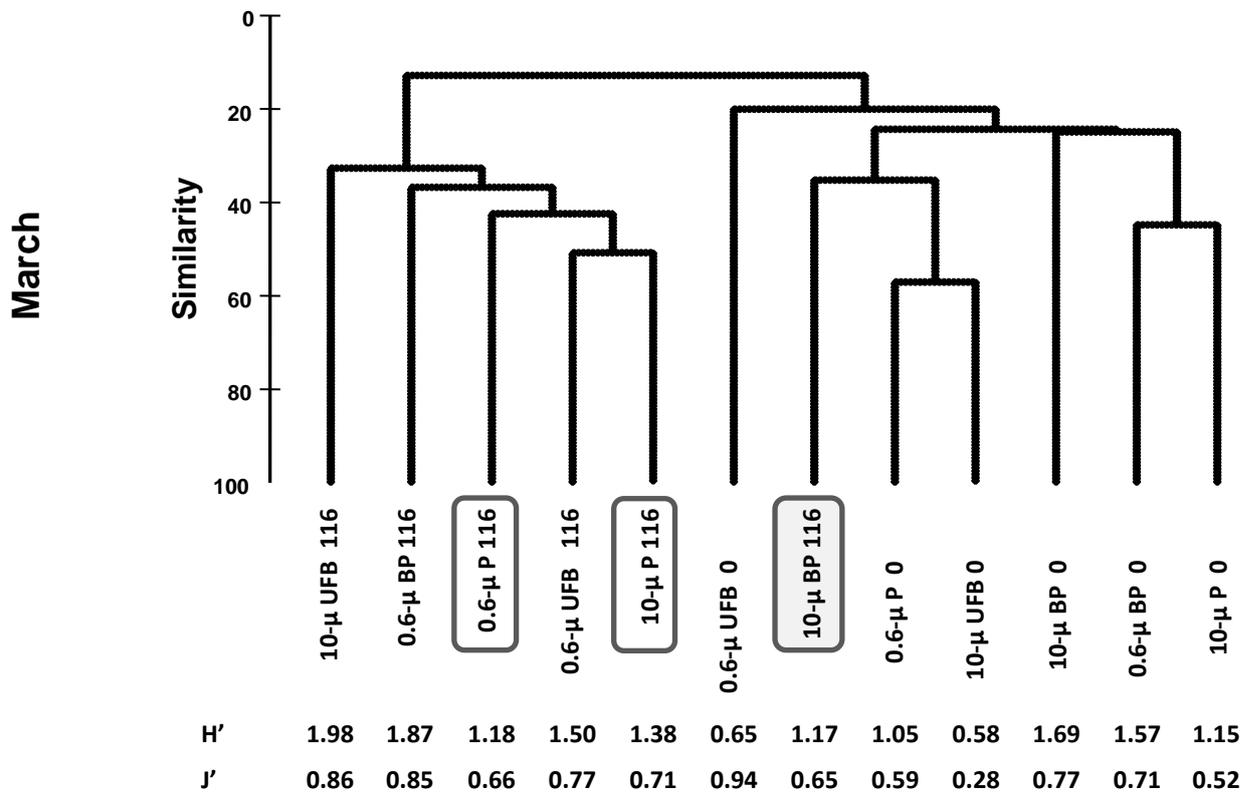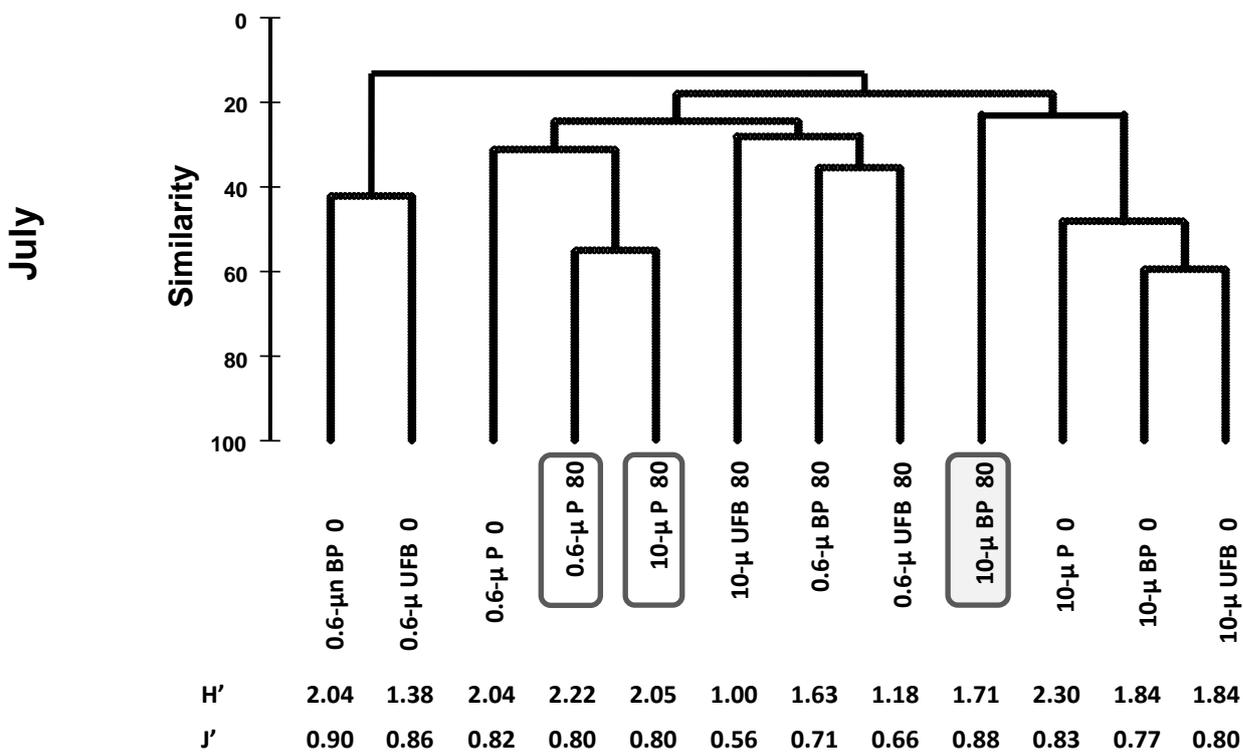

Fig. 5

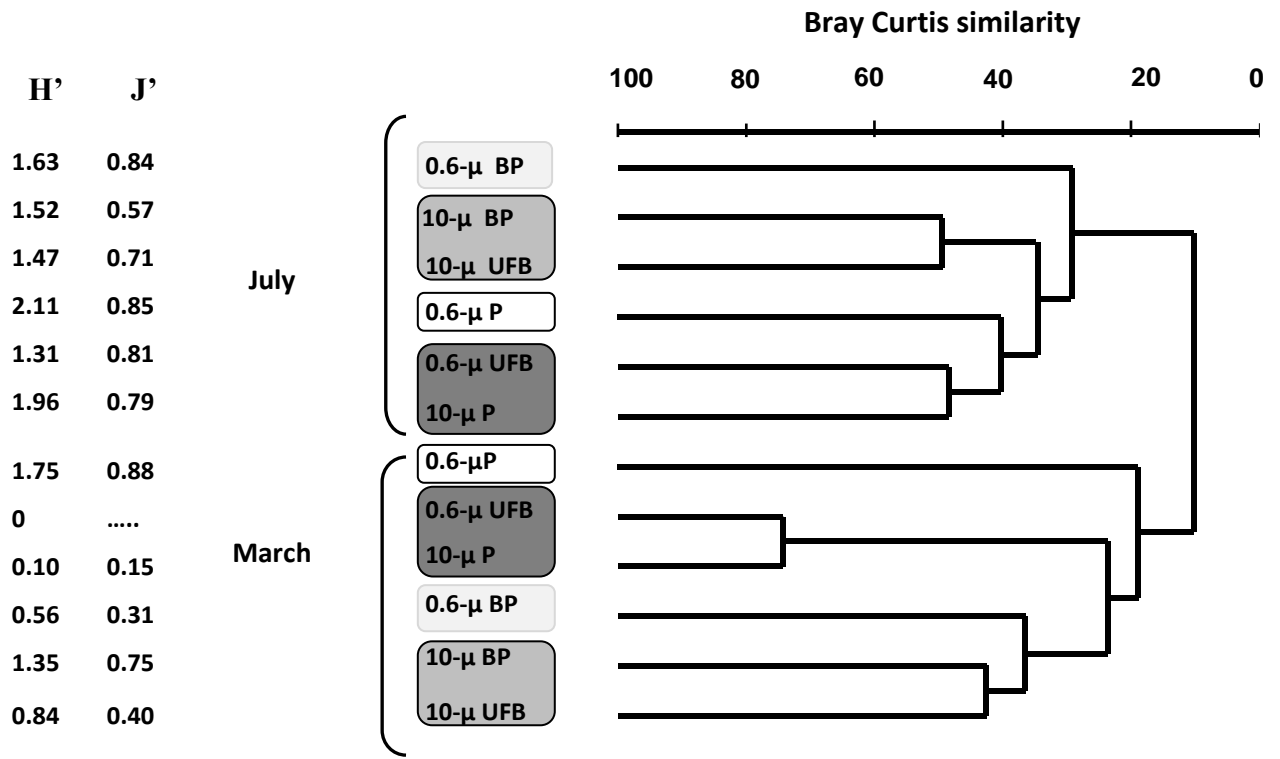

Fig. 6

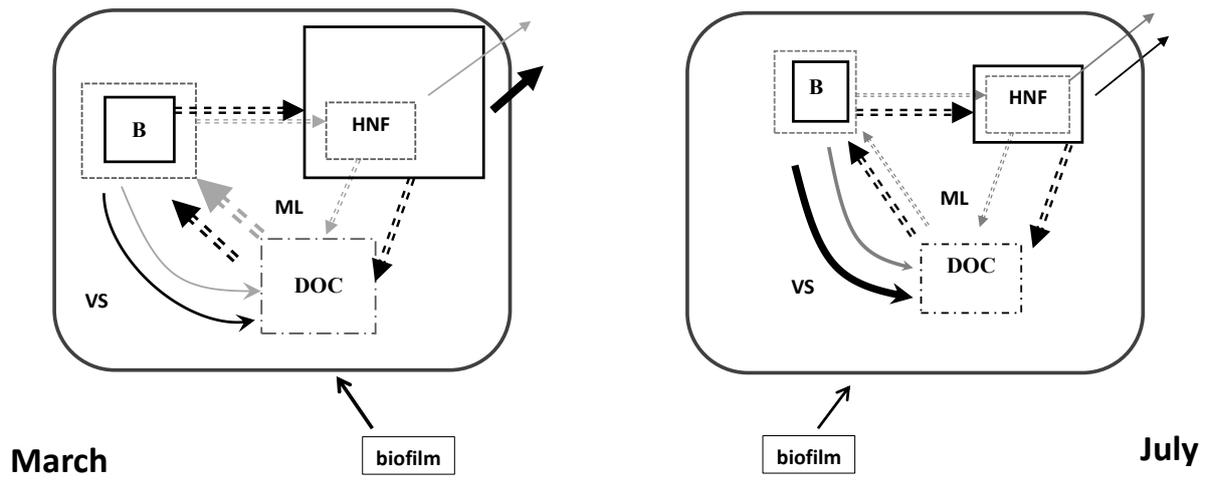

**Fig. 7**

Table 1: Significance of two-factors Model I ANOVA on the mean bacterial production (MBP), mean viral production (MVP) and the maximal net increase of bacteria (maxY, maximum yield at the plateau) for experimental data of March and July 2008. Nanoflagellate (HNF) and treatment factors and their interactions were tested with the F statistic. *A posteriori* multiple comparisons were made by a Bonferroni test with three treatment levels: controls (0.6-µ-P and 10-µ-P treatments) *vs.* benthic low molecular weight compounds LMWC (0.6-µ-UFB and 10-µ-UFB treatments) and resuspended biofilm (0.6-µ-BP and 10-µ-BP treatments).
*p* significance: * < 0.05, ** < 0.01, *** < 0.001, **** < 0.0001.

|  |  |  | HNF | Interaction | HNF effect | treatment effect | LMWC effect | biofilm effect |
|---|---|---|---|---|---|---|---|---|
| **March** | Bacteria | MBP | - | ns<br>p=0.4 | ***<br>p=0.0003 | ****<br>p<0.0001 | *** | ** |
|  |  |  | + |  |  |  | *** | ns |
|  |  | max-Y | - | **<br>p= 0.0048 | ns<br>p=0.46 | ***<br>p=0.0003 | *** | ns |
|  |  |  | + |  |  |  | ns | ns |
|  | Virus | MVP | - | ns<br>p=0.33 | ns<br>p=0.28 | ns<br>p=0.054 | ns | * |
|  |  |  | + |  |  |  | ns | ns |
| **July** | Bacteria | MBP | - | ns | ns<br>p=0.95 | *<br>p=0.011 | ns | ns |
|  |  |  | + |  |  |  | * | ns |
|  |  | max-Y | - | *<br>p=0.0102 | ***<br>p=0.0006 | ****<br>p<0.0001 | *** | ** |
|  |  |  | + |  |  |  | ** | ns |
|  | Virus | MVP | - | ns | *<br>p=0.012 | ns | ns | ns |
|  |  |  | + |  |  |  | ns | ns |